\documentclass[acus]{JAC2003}

\usepackage{float}
\usepackage{graphicx}

\begin{document}
\title{CONVERSION OF A TRANSVERSE DENSITY MODULATION INTO \\
A LONGITUDINAL PHASE SPACE MODULATION USING \\
AN EMITTANCE EXCHANGE TECHNIQUE \thanks{This work was supported by the Fermi Research Alliance, LLC under Contract No. DE-AC02-07CH11359 with the U.S. Department of Energy. P.P. was partially supported by the US Department of Energy under Contract No. DE-FG02-08ER41532  with Northern Illinois University.} }

\author{Y.-E Sun$^1$, P. Piot$^{1,2}$, A. Johnson$^{3}$, A. Lumpkin$^3$, J. Ruan$^3$, and R. Thurman-Keup$^3$ \\
$^1$ Accelerator Physics Center, Fermi National Accelerator Laboratory, Batavia, IL 60510, USA \\
$^2$ Department of Physics, Northern Illinois University DeKalb, IL 60115, USA \\
$^3$ Accelerator Division, Fermi National Accelerator Laboratory, Batavia, IL 60510, USA }

\maketitle

\begin{abstract}
We report on an experiment to produce a train of sub-picosecond microbunches using a transverse-to-longitudinal emittance exchange technique. The generation of a modulation on the longitudinal phase space is done by converting an initial horizontal modulation produced using a multislits mask.  The preliminary experimental data clearly demonstrate the conversion process. To date only the final energy modulation has been measured. However numerical simulations, in qualitative agreement with the measurements, indicate that the conversion process should also introduce a temporal modulation.
\end{abstract}
\begin{figure*}[tttthhhhh!!!!!!!!!]
\centering
\includegraphics[width=0.99\textwidth]{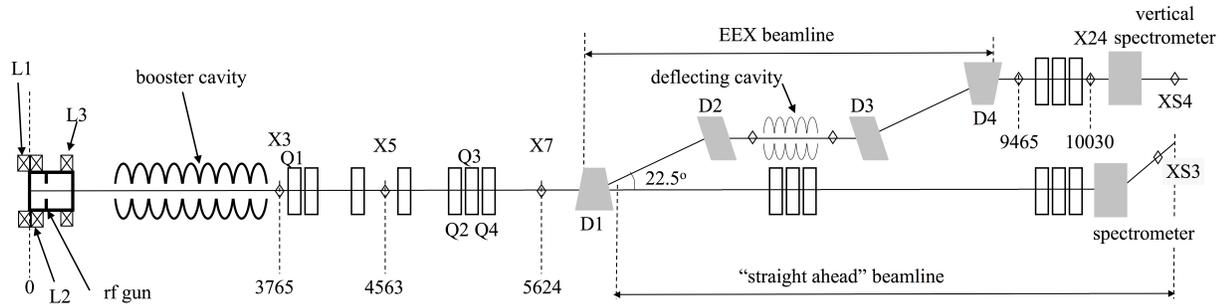}
\caption{Top view of the A0 photoinjector showing elements pertinent to the present experiment. The ``X" refers to diagnostic stations (beam viewers and/or multi-slit masks location), ``L" to the solenoidal lenses, ``Q" to quadrupoles. Distances are in millimeters. The spectrometer downstream of the emittance exchange beamline bends the beam in the vertical direction.}\label{fig:beamline}
\end{figure*}

\section{Introduction}
Recent years have witnessed an increasing demand for precise phase-space control schemes.
In particular, electron bunches with a well-defined temporal distribution are often desired. An interesting class of temporal distribution consists of train of microbunches with sub-picosecond duration and separation. Applications of such trains of microbunches include the generation of super-radiant radiation~\cite{gover} or the resonant excitation of wakefield in plasma wakefields accelerators or dielectric-loaded structures~\cite{jing}. To date there are very few techniques capable of reliably providing this class of beam~\cite{muggli}. We have recently explored an alternative technique based on the use of a transverse-to-longitudinal phase space exchange method~\cite{piotAAC08,yineLINAC08}.  The method consists of shaping the beam's transverse density to produce the desired horizontal profile. The horizontal profile is then mapped onto the longitudinal profile by a beamline capable of exchanging the phase space coordinates between the horizontal and longitudinal degrees of freedom. Therefore the production of a train of microbunches simply relies on generating a set of horizontally-separated beamlets upstream of the beamline.
The backbone of the proposed technique is the transverse-to-longitudinal phase space exchange which was recently proposed as a means to mitigate the microbunching instability in high-brightness electron beams~\cite{emma} or to improve the performance of single-pass FELs~\cite{emma2}. A simple optical lattice capable of performing this phase space exchange consists of a horizontally-deflecting cavity, operating on the TM$_{110}$ mode, flanked by two horizontally-dispersive sections henceforth refereed to as ``doglegs"~\cite{kim}. Under the thin lens approximation the initial transverse phase space coordinates $(x_0, x'_0)$ are mapped to the longitudinal phase space coordinates $(z,\delta)$ following~\cite{yine}
\begin{eqnarray}
\left\{ \begin {array}{ll} z =& -\frac{\xi}{\eta}x_0-\frac{L\xi-\eta^2}{\eta}x_0'\\
\delta =& -\frac{1}{\eta}x_0-\frac{L}{\eta}x_0',
\end{array}\right.
\end{eqnarray}
where $L$ is the distance between the dogleg's dipoles, and $\eta$ and $\xi$ are respectively the horizontal and longitudinal dispersions generated by one dogleg. The coupling described by Eq.~1 can used to arbitrarily shape the temporal distribution of an electron beam~\cite{piotPRSTAB}.

\section{Experimental setup}
The experiment was carried out at the Fermilab's A0 Photoinjector~\cite{carneiro}; see Fig.~\ref{fig:beamline}. In brief, electron bunches are generated via photoemission from a cesium telluride photocathode located on the back plate of a 1+1/2 cell radio-frequency (rf) cavity operating at 1.3~GHz (the ``rf gun"). The rf gun is surrounded by three solenoidal lenses that control the beam's transverse size and emittance. The beam is then accelerated in a 1.3~GHz superconducting rf cavity (the booster cavity) to approximately 16~MeV. Downstream of the booster cavity, the beam line includes a set of quadrupoles, steering dipoles, and diagnostics before splitting into two lines.

The ``straight ahead" beamline incorporates an horizontally-bending spectrometer equipped with a yttrium aluminum garnet (YAG) screen (labeled XS3 in Fig.~\ref{fig:beamline}) for energy measurement. The horizontal dispersion value at XS3 location is $|\eta_{x,XS3}| =317$~mm.

The other beamline, the emittance exchange (EEX) beamline, implements the aforedescribed double-dogleg emittance exchanger~\cite{koeth0}. The doglegs consist of dipole magnets with $\pm 22.5^{\circ}$ bending angles and each generate a horizontal and longitudinal dispersions of $\eta\simeq -33$~cm and $\xi\simeq -12$~cm, respectively~\footnote{In our convention the head of the bunch is for $z<0$.}.  The deflecting cavity is  a liquid-Nitrogen--cooled, normal-conducting, five-cell cavity operating on the TM$_{110}$ $\pi$-mode at 3.9~GHz~\cite{koeth}. The lattice downstream of the EEX beamline includes three quadrupoles, a suite of standard beam diagnostics and a vertical spectrometer. At X24, the total power emitted via coherent transition radiation (CTR) as the beam impinges an Aluminum screen can be measured using a pyroelectric detector. Such a measurement can be used to minimize the bunch length at X24 by maximizing the radiated power. The vertical dispersion generated by the spectrometer at the XS4 YAG screen is  $ |\eta_{y,XS4}|= 944 $~mm. The nominal operating parameters relevant to the experiment reported below are gathered in Table~\ref{tab:operating}.

\begin{table}[h!]
\caption{\label{tab:operating} Nominal settings for the rf-gun,
booster cavity, and the photo-cathode drive-laser.}
\begin{center}
\begin{tabular}{l c c}
\hline \hline parameter                       &      value       &
units  \\ \hline
laser injection phase$^a$           &  45 $\pm$ 5     & rf deg \\
laser radius on cathode         & 0.9    & mm     \\
laser rms pulse duration            &  3.0 $\pm$ 0.5  & ps     \\
bunch charge$^b$                    &  $360\pm 20$    & pC \\
$E_z$ on cathode                &  33.7  $\pm$ 0.2  & MV/m     \\
peak $B_z$$^c$ (L2, L3)       &  (0.158, 0.041)   & T     \\
booster cavity acc. field    &  $\sim$ 11.5    & MV/m     \\
booster cavity phase$^d$    &  $\sim$ 25    & rf deg     \\
\hline \hline
\end{tabular}
\end{center}
$^a$ {\small the phase is referenced w.r.t the zero-crossing phase,}\\
$^b$ {\small charge before interception by  the X3 slits,}\\
$^c$  {\small the peak field of solenoid L1 was tuned to zero the axial magnetic field on the photocathode,} \\
$^d$  {\small the phase is referenced w.r.t. the maximum energy and corresponds to a minimum fractional momentum spread. }
\end{table}

\section{experimental methods \\
and results}

For the proof-of-principle experiment, the production of a transverse modulation was achieved by passing the beam through a set of remotely insertable vertical slits at X3. This multislit mask, nominally designed for single-shot transverse emittance measurements, consists of 50~$\mu$m wide slits made out of a 3~mm thick tungsten plate. The slits are separated by 1~mm. Less than 5~\% of the incoming beam is transmitted through the mask. This mask is obviously not optimized for the present experiment: a smaller slits spacing would be more beneficial to increase the total transmitted charge and to get a smaller separation between the bunches after the EEX beamline.

\begin{figure}[hhhhhhhhh!!!!!!!!!!!!!!]
\includegraphics[scale = 0.49]{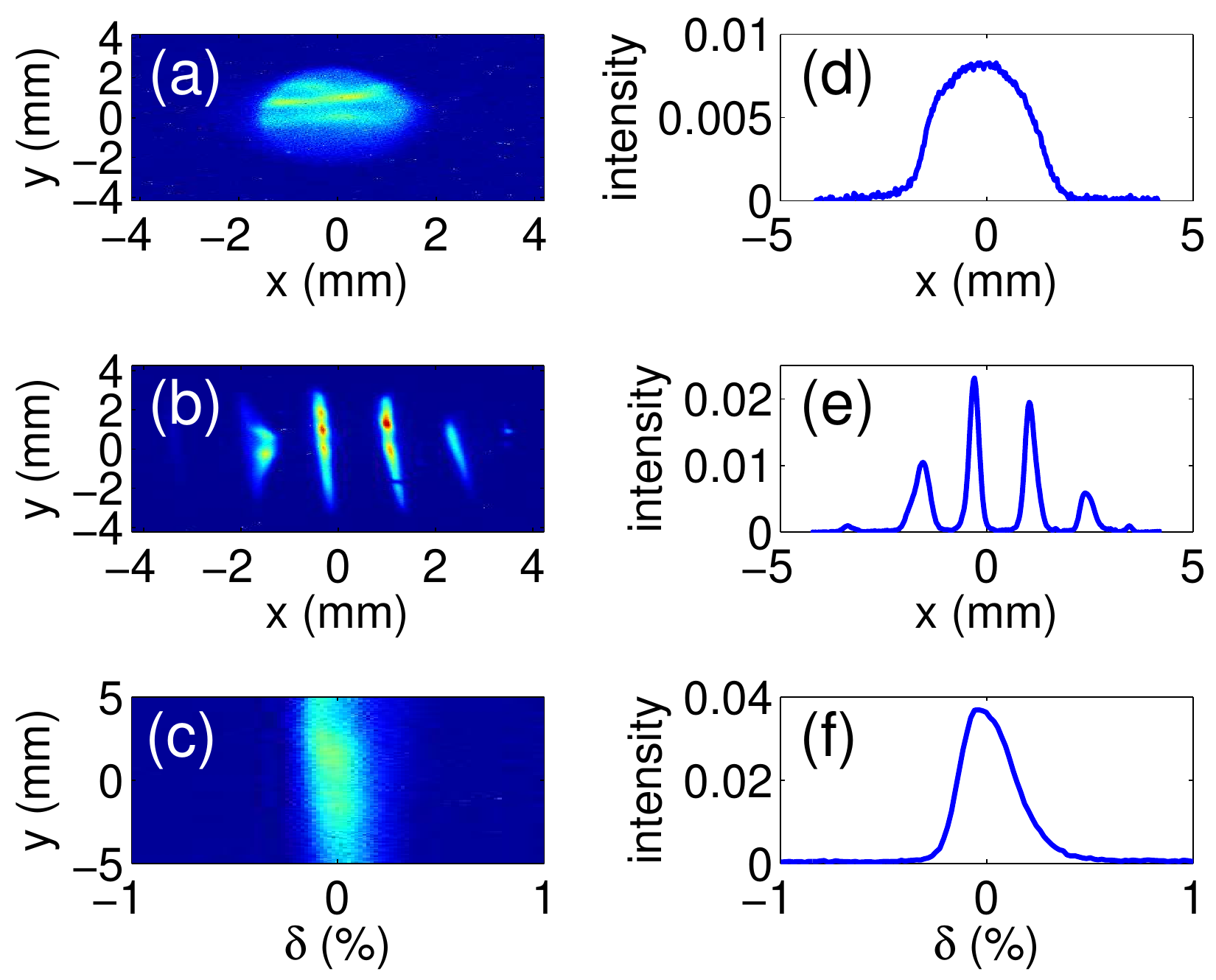}
\caption{Characterization of the transversely modulated beam upstream of the emittance exchanger showing the beam transverse density at X3 (before inserting the multislit mask) (a), the beam density at X5 after insertion of the multislit mask at X3 (b), an example of beam energy spectrum measured at XS3 with X3 slits inserted (c). The plots (d, e, f) in the right column are the corresponding horizontal projections. \label{fig:initdistribution}}
\end{figure}

The beam was first diagnosed in the straight-ahead line and the spectrometer to ensure especially that no energy modulation was observable. The resulting set of observations is depicted in Fig.~\ref{fig:initdistribution}.

\begin{figure}[hhhhhhhhh!!!!!!!!!!!!!!]
\includegraphics[scale = 0.45]{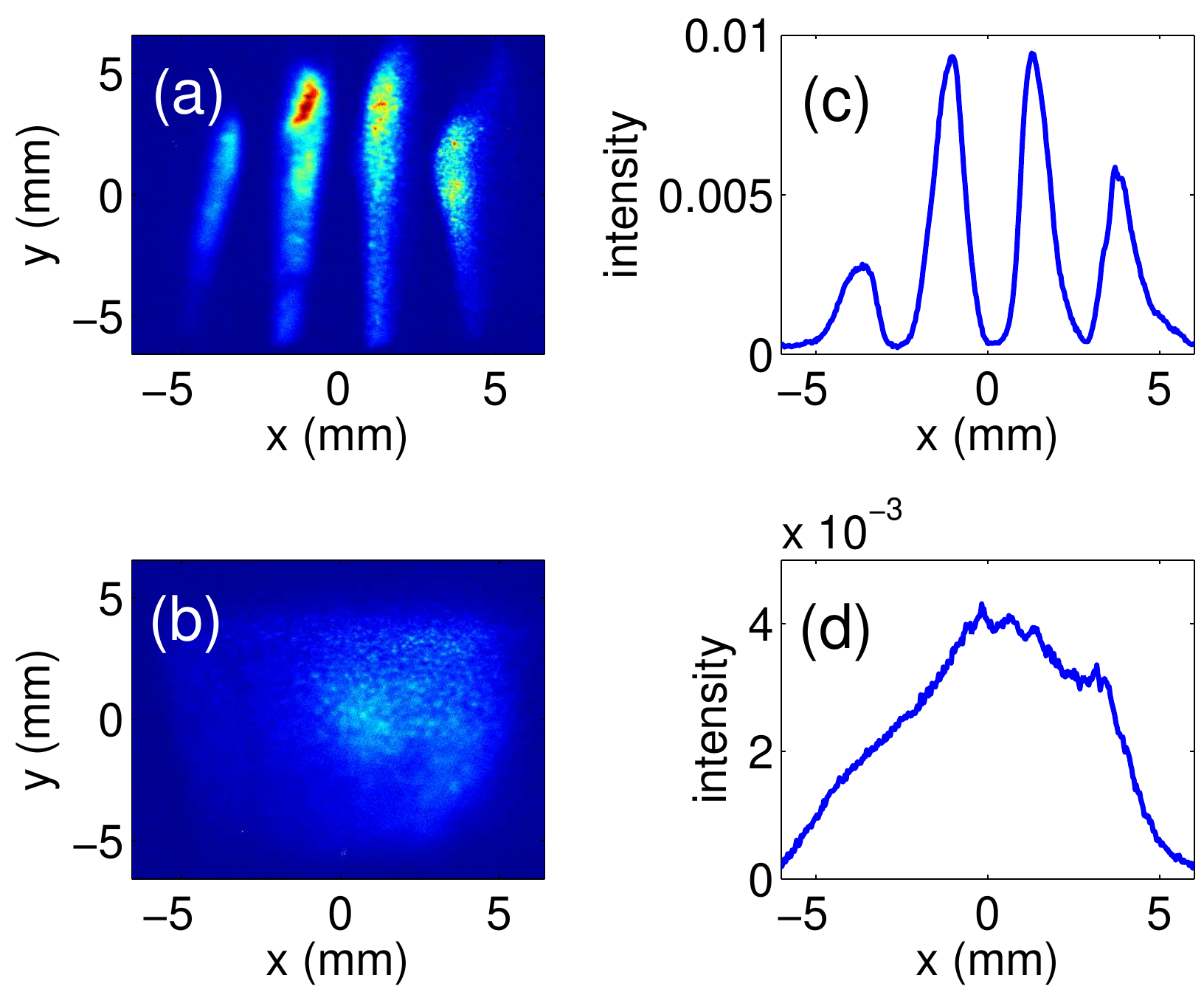}
\caption{Characterization of the transversely modulated beam downstream of the emittance exchanger showing the beam transverse density at X24 with the deflecting cavity off (a) and on (b), The plots (c,d) in the right column are the corresponding horizontal projections. \label{fig:afterEEX}}
\end{figure}

\begin{figure}[hhhhhhhhh!!!!!!!!!!!!!!]
\includegraphics[scale = 0.44]{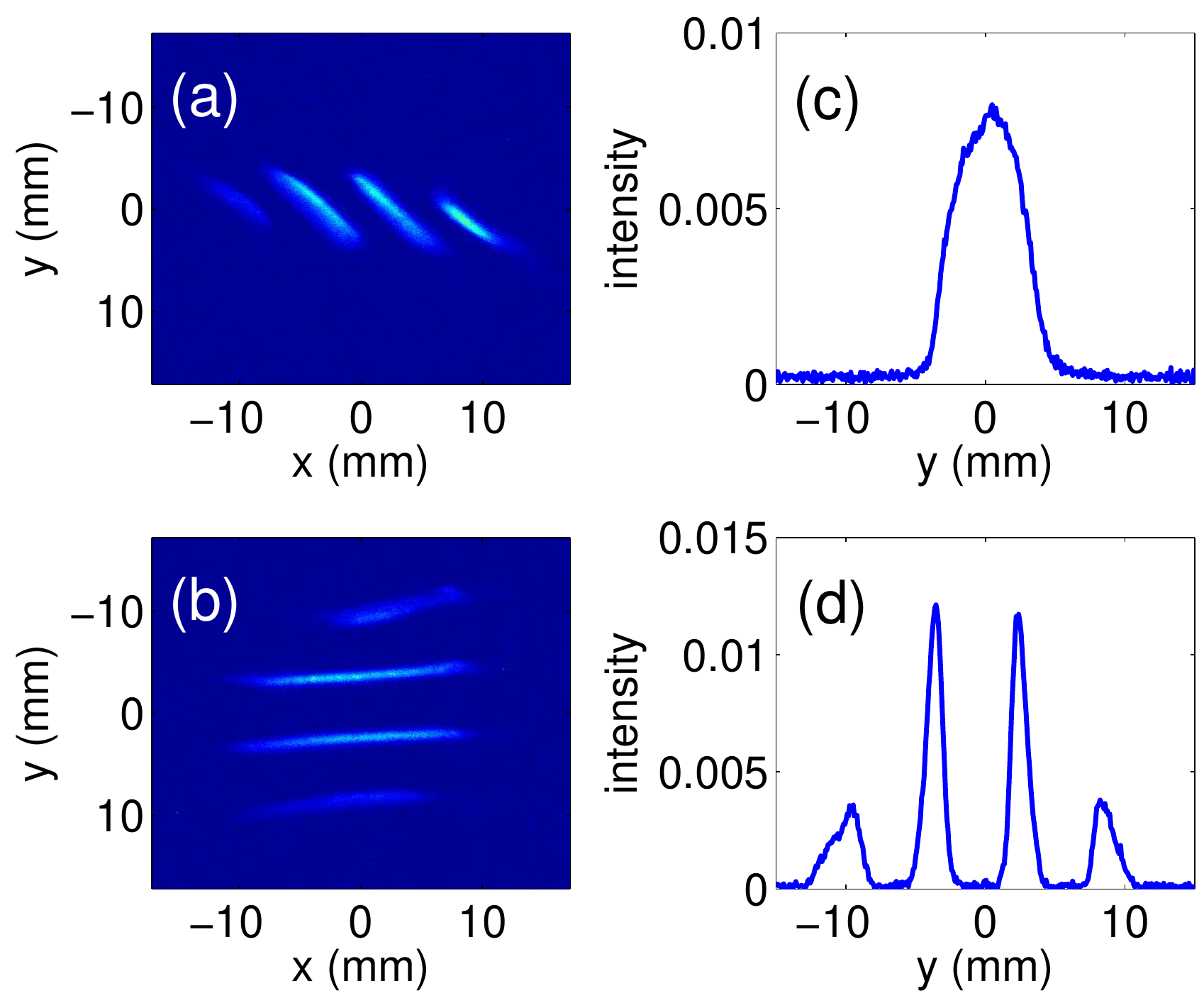}
\caption{Characterization of the transversely modulated beam upstream of the emittance exchanger showing the beam transverse density at XS4 with the deflecting cavity off (a) and on (b), The plots (c,d) in the right column are the corresponding vertical projections. The vertical axis $y$ is proportional to the fractional momentum spread. \label{fig:XS4EEX}}
\end{figure}

The beam was then transported through the EEX beamline with the deflecting cavity turned off. The transverse modulation was still observable at X24 but no energy modulation could be seen at XS4. Powering the cavity to its nominal deflecting voltage ($V_{x}\simeq 290$~kV) resulted in the suppression of the transverse modulation at X24 and the appearance of an energy modulation at XS4; see Fig.~\ref{fig:afterEEX} and Fig.~\ref{fig:XS4EEX}. These observations clearly demonstrate the ability of the EEX beamline to convert an incoming transverse density modulation into an energy modulation. In the present measurement the incoming horizontal Courant-Snyder (CS) parameters at the EEX beamline entrance were empirically tuned to maximize the energy modulation. However other C-S parameters settings can map the initial modulation into a temporal modulation as can be inferred from Eq.~1.

\section{numerical simulations}
Start-to-end simulations were performed in order to provide insight on the experimental results. The program {\sc astra}~\cite{astra} was used to model the beam dynamics from the cathode to X3.  The effect of the multislit mask was simulated and the resulting distribution was tracked up to XS4 using  {\sc elegant}~\cite{elegant} without including collective effects.  Space charge effects are not expected to be significant in driving the evolution of the beamlet dynamics as can be inferred by estimating the ratio of the space charge and emittance terms in the envelope equation ${\cal R} =\hat{I}/[3\sqrt{2\pi} I_A \gamma](w/\varepsilon_x)^2$ where $\hat I\simeq32$~A is the peak current (we take the temporal distribution to be Gaussian with rms duration $\sigma_t=4.5$~ps), $\gamma$ is the beam's Lorentz factor, $w=50$~$\mu$m is the slits width, and $I_A=17$~kA is the Alf\`en current. For the considered beam parameters we found ${\cal R}\simeq 8\times 10^{-3} \ll 1$ supporting the use of {\sc elegant}.

\begin{figure}[hhhhhhhhh!!!!!!!!!!!!!!]
\includegraphics[scale = 0.45]{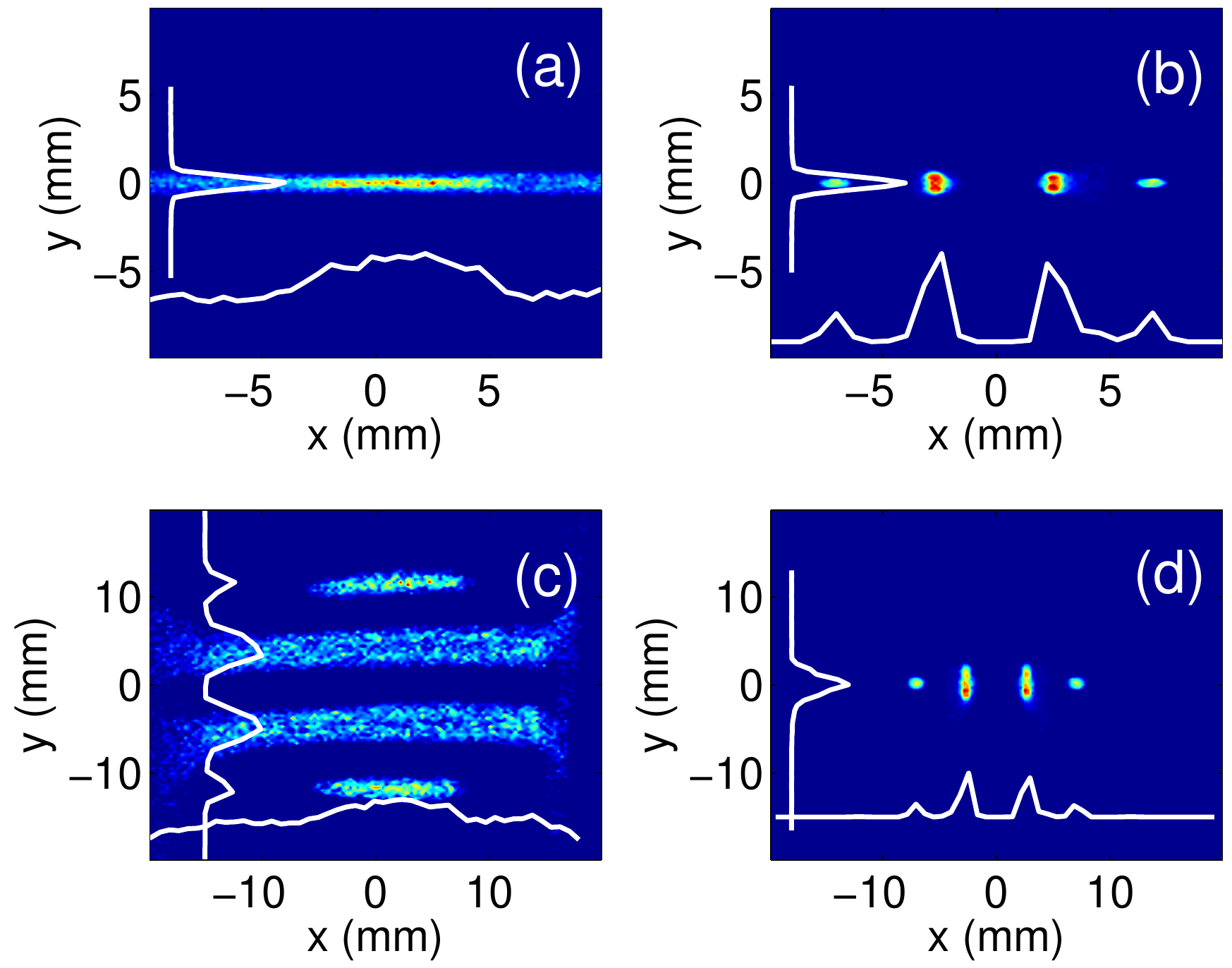}
\caption{Simulated density distribution at X24 and XS4 (respectively top and bottom rows) for transverse deflecting cavity on [left column images (a) and (c)] and off [right column, images (b) and (d)]. In images (c) and (d) the vertical axis $y$ is proportional to the fractional momentum spread.\label{fig:XS4simu}}
\end{figure}

The accelerator settings gathered in Table~\ref{tab:operating} were used in {\sc astra} and {\sc elegant} as a starting point for our simulations. In {\sc elegant} the settings of quadrupoles Q1, Q2, Q3 and Q4 were altered per the operating procedure followed during the experiment. First, the quadrupoles were scanned to minimize the final bunch length and energy spread without inserting X3 slits. The quadrupoles were then empirically tuned to maximize the energy separation of the beamlet on XS4. The resulting images at X24 and XS4 with deflecting cavity on and off are shown in Fig.~\ref{fig:XS4simu}.  Although there is a qualitative agreement between simulations and measurements, further improvement of the numerical model is needed before confidently benchmarking simulations and experiments.

\begin{figure}[hhhhhhhhh!!!!!!!!!!!!!!]
\includegraphics[scale = 0.47]{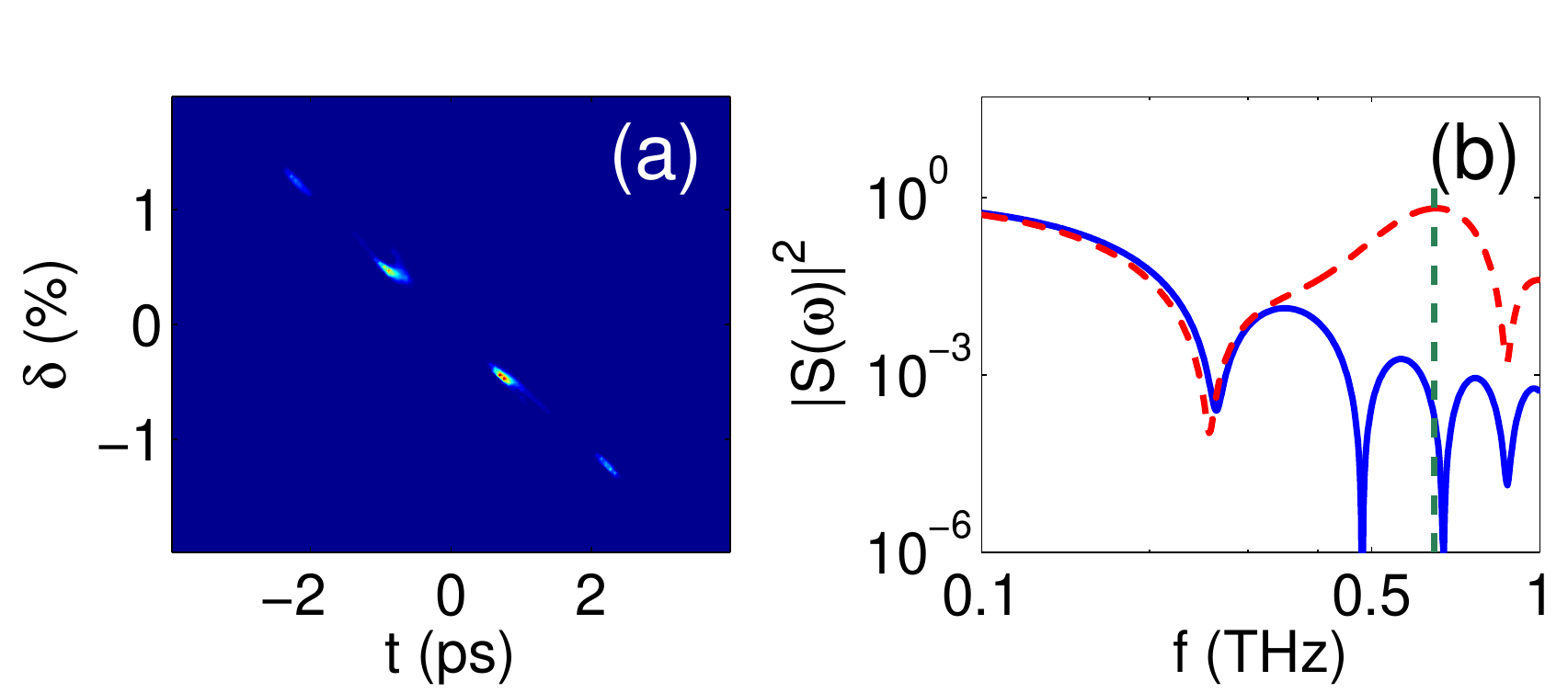}
\caption{Simulated longitudinal phase space $(t,\delta)$ at X24 (a) and resulting bunch form factor [red trace in plot (b)]. The simulations parameters are identical to those used to generate images (a) and (c) in Fig.~\ref{fig:XS4simu}. The blue trace in plot (b) is the bunch form factor computed for the entire beam (360 pC) at X24 i.e. when the X3 slits are not inserted. In image (a) $t>0$ corresponds to the tail of the bunch. \label{fig:tdpp}}
\end{figure}

The corresponding longitudinal phase space downstream of the EEX beamline is shown in Fig.~\ref{fig:tdpp}. The phase space is significantly correlated and therefore has both an energy and temporal modulations.

The spectral intensity radiated by a bunch of $N$ electrons by the CTR process is related to the single electron intensity  $\frac{dI}{d\omega}\big|_1$ via $\frac{dI}{d\omega }\big|_N=\frac{dI}{d\omega }\big|_1 [N+N^2|S(\omega)|^2]$ where $S(\omega)$ is the intensity-normalized Fourier transform of the normalized charge distribution $S(t)$~\cite{saxon}. Considering a series of $N_b$ identical microbunches  with normalized distribution $\Lambda(t)$  we have $S(t)=N_b^{-1} \sum_{n=1}^{N_b} \Lambda(t+nT)$ (where $T$ is the period) giving  $|S(\omega)|^2 =  \Xi  |\Lambda(\omega)|^2$. The intra-bunch coherence factor  $\Xi= N_b^{-2} \sin^2(\omega N_b T/2)/[\sin^2(\omega T/2)]$ describes the enhancement of radiation emission at resonance, i.e. for frequencies $\omega_n = 2\pi n /T$.  The computed bunch form factors (BFFs) $|S(\omega)|^2$ from the simulated macroparticle ensemble are shown in Fig.~\ref{fig:tdpp} for the case of the entire bunch, i.e. without slits X3 inserted, (blue solid trace) and for the case corresponding to image (a) of Fig.~\ref{fig:tdpp} (red dash trace). At $f\simeq 0.7$~THz [indicated by the vertical green dash line in Fig.~\ref{fig:tdpp} plot (b)] the microbunch train BFF is strongly enhanced due to the intra-bunch coherence and is higher than the single bunch BFF.

\section{Summary and outlook}
In summary we have shown that an incoming transverse modulation could be converted into the longitudinal phase space using the EEX beamline. In the preliminary set of experimental results  reported in this paper, the accelerator parameters were empirically chosen to obtain a clear energy modulation. The data are qualitatively consistent with single-particle tracking which indicate that a temporal modulation is also present due to the strongly correlated longitudinal phase space. In the near future we plan on demonstrating that the technique can produce a train of sub-picosecond microbunches by directly measuring the temporal distribution. The train of microbunches will be characterized via autocorrelation of the CTR radiation emitted at X24.  Because of the low bunch charge ($Q\simeq 10-15$~pC after the X3 slits) a Helium-cooled bolometer will be used to detect CTR. 

\section{Acknowledgments}
We are indebted to W. Muranyi, J. Santucci, and B. Tennis for their excellent operational and technical supports. We thank M. Church, H. Edwards, E. Harms and V. Shiltsev for their interest and encouragement.

\end{document}